\documentclass[a4paper,11pt]{article}
\usepackage{pos}
\usepackage{float}  
\usepackage{subcaption}

\title{Hadronic light-by-light contribution to the muon g-2 using twisted-mass fermions}

\author*[a]{N.~Kalntis}
\author[a,b]{G.~Kanwar}
\author[c]{M.~Petschlies}
\author[a]{S.~Romiti}
\author[a]{U.~Wenger}

\affiliation[a]{Institute for Theoretical Physics, Albert Einstein Center for Fundamental Physics, University of Bern, CH-3012 Bern, Switzerland}
\affiliation[b]{Higgs Centre for Theoretical Physics, University of Edinburgh, Edinburgh EH9 3FD, UK}
\affiliation[c]{Helmholtz-Institut f{\"u}r Strahlen- und Kernphysik (Theorie), University of Bonn, 53115 Bonn, Germany}
\emailAdd{nkalntis@itp.unibe.ch}

\abstract{We report on our preliminary results from the lattice-QCD computation of the hadronic light-by-light (HLbL) contribution to the anomalous magnetic moment of the muon. We use twisted-mass $N_f = 2 + 1 + 1$ gauge ensembles at the physical point
generated by the Extended Twisted Mass Collaboration (ETMC).
\begin{center}
\includegraphics[width=0.3\textwidth]{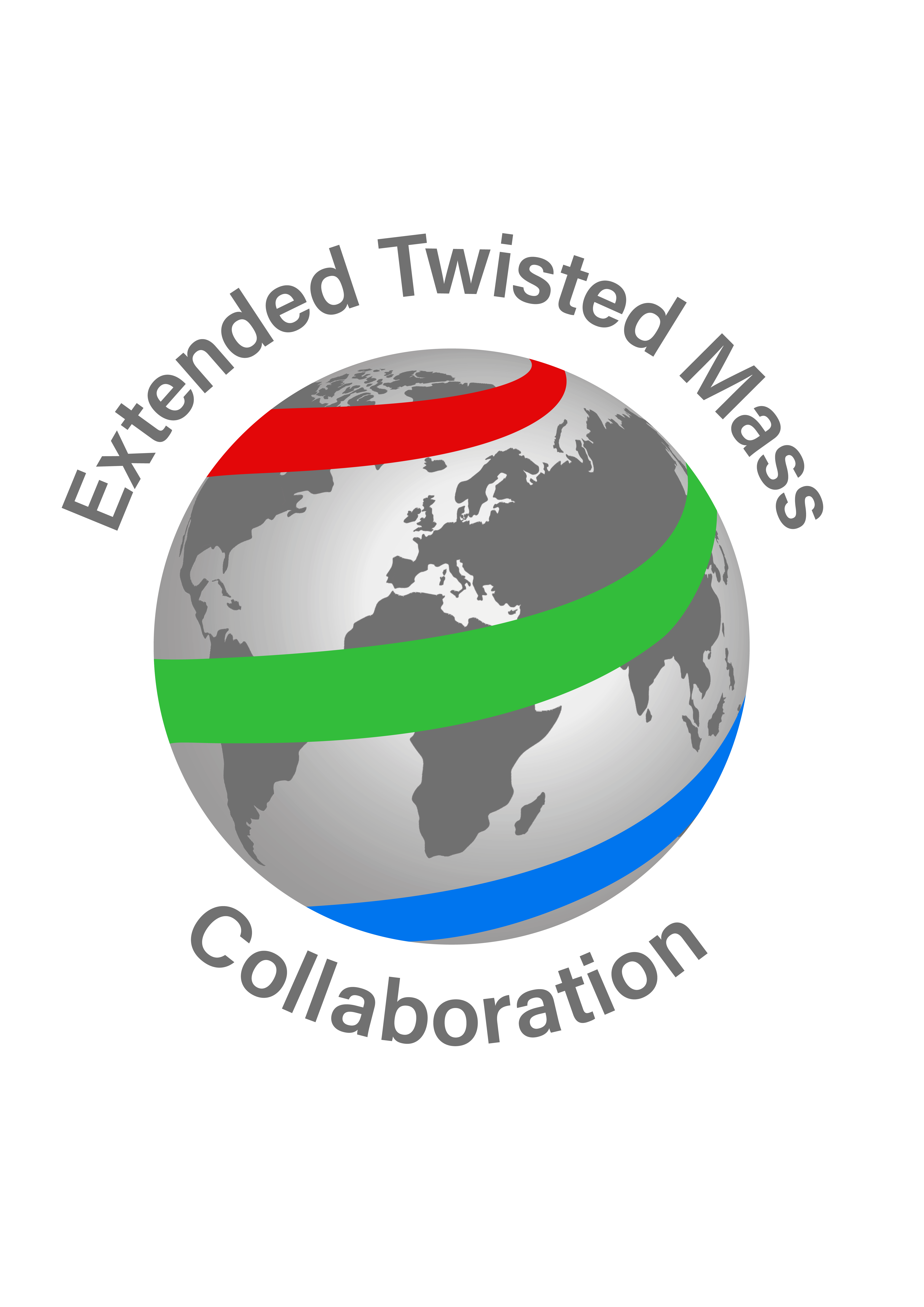}
\end{center}
}

\FullConference{The 41st International Symposium on Lattice Field Theory (LATTICE2024)\\
 28 July - 3 August 2024\\
Liverpool, UK\\}

\begin{document}
\maketitle

\section{Introduction}

We compute the hadronic light-by-light (HLbL) contribution to the anomalous magnetic moment of the muon from lattice QCD. For this computation we use twisted-mass clover-improved fermions at maximal twist, which leads to automatic $\mathcal{O}(\text{a})$ improvement on the calculated observables \cite{Frezzotti:2003ni, Frezzotti:2004wz}. We use three $N_f = 2+1+1$ gauge ensembles generated by the Extended Twisted Mass Collaboration (ETMC) \cite{ExtendedTwistedMass:2021gbo,ExtendedTwistedMass:2022jpw}. The quark masses for these ensembles are tuned such that the charged-pion mass is fixed to its physical value and
the $s$- and $c$-quark masses are tuned to reproduce the physical $\Omega$- and $\Lambda_c$-baryon masses. The details of these ensembles are summarized in Table \ref{tab:ensemble_details}.

\begin{table}[h!]
\small
\centering
\[
\begin{array}{|c||c|c|c|c|c|c|c|}
\hline
\text{Ensemble} & L^3 \cdot T /a^4 & M_{\pi}~[\text{MeV}] & a~[\text{fm}] & L~ [\text{fm}] & M_{\pi} \cdot L & Z_V & Z_A \\ \hline\hline
\text{cB64} & 64^3 \cdot 128 & 140.2(2) & 0.07961(13) & 5.09 & 3.62 & 0.706379(24)  & 0.74294(24) \\ \hline
\text{cC80} & 80^3 \cdot 160 & 136.7(2) & 0.06821(12) & 5.46 & 3.78 & 0.725404(19) & 0.75830(16) \\ \hline
\text{cD96} & 96^3 \cdot 192 & 140.8(2) & 0.05692(10) & 5.46 & 3.90 & 0.744108(12)  & 0.77395(12) \\ \hline
\end{array}
\]
\caption{Details of the ensembles used for the calculation of the HLbL contributions reported in these proceedings.}
\label{tab:ensemble_details}
\end{table}


We follow the position-space approach put forward by the Mainz collaboration \cite{Green:2015mva, Asmussen:2016lse, Chao:2020kwq, Chao:2021tvp,Chao:2022xzg}, where the HLbL contribution to $a_\mu$ can be calculated as 

\begin{equation} 
    a_\mu^{\text{HLbL}} = \frac{me^6}{3} \int d^4 x \, d^4 y~\Bar{\mathcal{L}}_{[\rho,\sigma];\mu\nu\lambda}(x,y)~i \Hat{\Pi}_{\rho;\mu\nu\lambda\sigma}(x,y) \, . 
    \label{eq:master_formula}
\end{equation}
In Eq.~\eqref{eq:master_formula}, the QED kernel $\Bar{\mathcal{L}}$ is associated with the perturbative QED part, which can be computed in the continuum and infinite volume, while the hadronic four-point  function $\Hat{\Pi}$ is calculated on a discrete lattice and in finite volume. Hence, working in position space enables the use of an infinite-volume QED kernel which describes the (long-distance) QED effects, while the hadronic contribution is treated fully non-perturbatively on the lattice. We note that the RBC-UKQCD collaboration \cite{Blum:2017cer,Blum:2019ugy,Blum:2023vlm} also uses a position-space approach, however with a different treatment of QED (both in finite and infinite volume). 

The four-point function is defined by
\begin{equation} 
    i \Hat{\Pi}_{\rho;\mu\nu\lambda\sigma}(x,y) = - \int d^4z~ z_\rho \langle j_\mu(x) j_\nu(y) j_\sigma(z) j_\lambda(0) 
    \rangle_{\text{QCD}} \,,
    \label{eq:Pi_hat}
\end{equation}
with $j_\mu(x)$ the electromagnetic current 
\begin{equation}
 j_\mu (x) = \sum_{f = u, d, s, c} Q_{f}~(\Bar{q}_f \gamma_{\mu} q_f) (x) \, 
\end{equation}
and $Q_f$ the electromagnetic charges of the various quark flavours. 
We remark that the nonconserved local currents do not spoil the automatic $\mathcal{O}(\text{a})$ improvement \cite{Frezzotti:2003ni, Frezzotti:2004wz} and  the four-point function in Eq.~(\ref{eq:Pi_hat}) is therefore $\mathcal{O}(\text{a})$ improved. The evaluation of the QCD four-point function contains five classes of Wick contractions, which are referred to as the (fully-)connected, and the $(2 + 2), (3 + 1), (2 + 1 + 1)$,  and $(1 + 1 + 1 + 1)$ disconnected contractions, cf.~Ref.~\cite{Chao:2020kwq}. In our computations we focus on the dominant contributions, which are the connected and the (2+2) ones.  
Using $O(4)$ rotational symmetry one can rewrite Eq.~\eqref{eq:master_formula} as $a_\mu^{\mathrm{HLbL}} = \displaystyle\lim_{|y|_{\mathrm{max}} \to \infty} a_\mu^{\text{HLbL}}(|y|_{\text{max}})$, where 

\begin{equation} \label{eq:amu-sum}
   a_\mu^{\mathrm{HLbL}}(|y|_{\mathrm{max}}) \equiv \int_0^{|y|_{\text{max}}} d |y| f(|y|)
\end{equation}
is the partially integrated contribution and
\begin{equation} \label{eq:f_y_final}
    f(|y|) = \frac{me^6}{3} 2 \pi^2 |y|^3  \int d^4x \Bar{\mathcal{L}}_{[\rho,\sigma];\mu\nu\lambda}(x,y)~i \Hat{\Pi}_{\rho;\mu\nu\lambda\sigma}(x,y) \,.     
\end{equation}
\noindent 
On the lattice, all spacetime integrals 
become sums over the lattice points. In our calculation, we perform a full summation over  $z$ in Eq.~\eqref{eq:Pi_hat} and $x$ in Eq.~\eqref{eq:f_y_final},
while we select specific points $y$ corresponding to a certain range of $|y|$ and integrate the one-dimensional integral in Eq.~\eqref{eq:amu-sum} by quadrature. The function $f(|y|)$ is referred to in the following as the integrand. We use the two modified kernels $\overline{\mathcal{L}}^{(3)}$ and $\overline{\mathcal{L}}^{(\Lambda=0.4)}$ as defined in Refs.~\cite{Asmussen:2019act, Chao:2020kwq}. These kernels integrate to the same value in the continuum and infinite-volume limit. However, on a discrete and finite lattice they lead to different discretization and finite-volume effects when contracted with the four-point function $\Hat{\Pi}$ and integrated (summed) over the finite lattice volume. Working with the two kernels can thus give an estimate of the lattice artefacts and the typical size of finite-volume effects. A thorough discussion on the analytic form of the various QED kernels can be found in Ref.~\cite{Asmussen:2022oql}. 

\section{Preliminary results}

Here we present some preliminary results for our calculations of the charm-, the strange- and the light-quark connected contributions, as well as the light-quark disconnected (2+2) contribution.

\subsection{Charm- and strange-quark connected contribution}
\label{subsec:charm_connected}

\begin{figure}[h!]
    \centering
    \begin{subfigure}[b]{\textwidth}
        \centering
        \includegraphics[width=\textwidth]{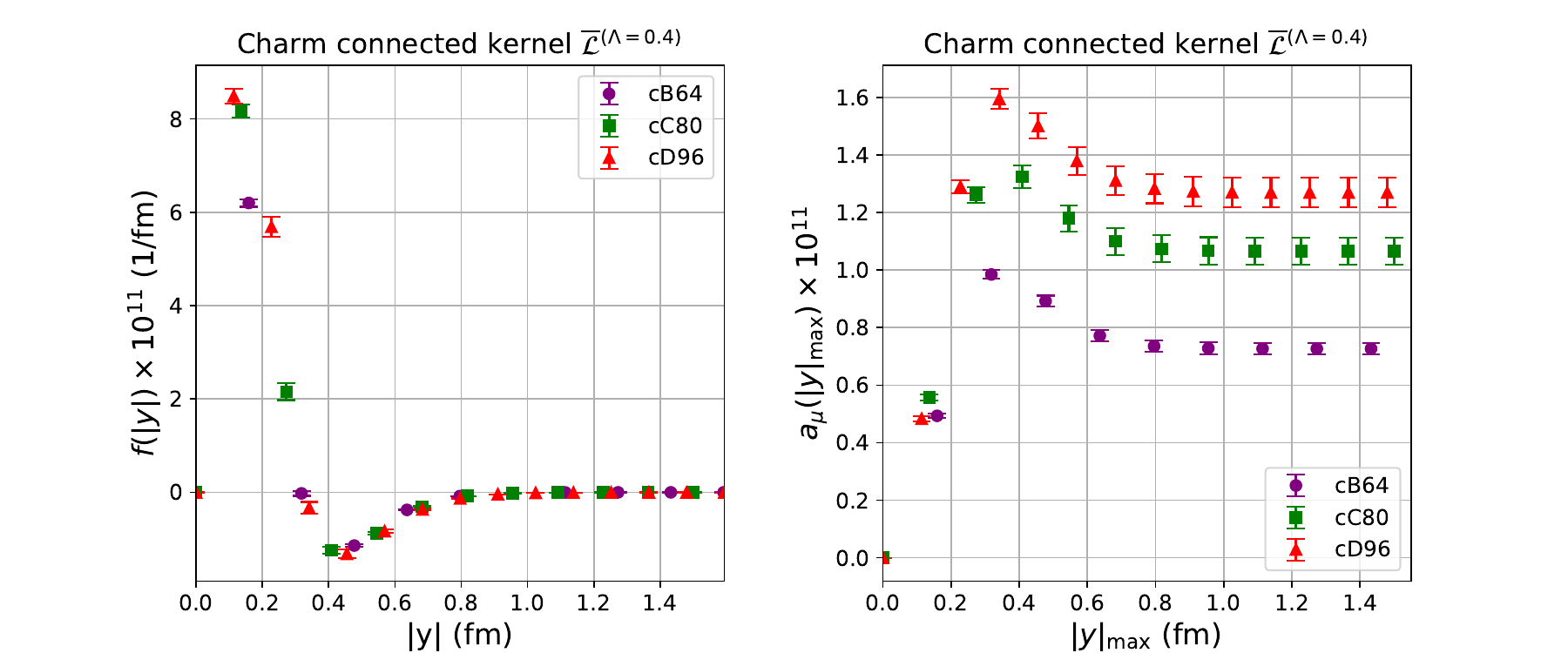}
        \label{fig:charm-k4}
    \end{subfigure}
    
    \begin{subfigure}[b]{\textwidth}
        \centering
        \includegraphics[width=\textwidth]{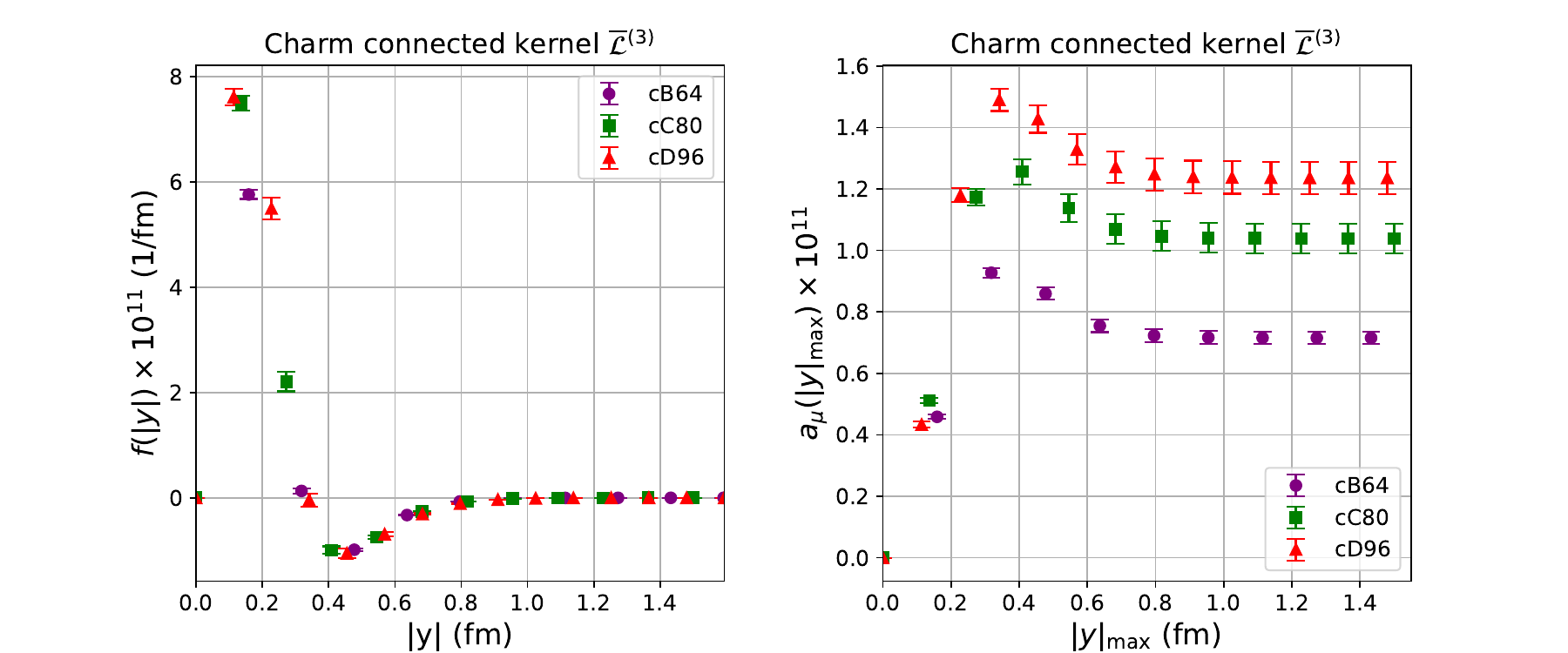}
        \label{fig:charm-k3}
    \end{subfigure}
    
    \caption{Preliminary results for the charm-quark connected contribution on the ensembles cB64, cC80, and cD96 for kernels $\overline{\mathcal{L}}^{(\Lambda=0.4)}$ (upper two plots) and $\overline{\mathcal{L}}^{(3)}$ (lower two plots). In both cases, the left plot depicts the integrand $f(|y|)$ as a function of the distance $|y|$ and the right plot the partially integrated $a_\mu(|y|_\text{max})$ as a function of the cut-off distance $|y|_\text{max}$.}
    \label{fig:charm-kernels}
\end{figure}

\begin{figure}[h!]
    \centering
    \hspace*{-0.5cm} 
    \includegraphics[width=0.8\textwidth]{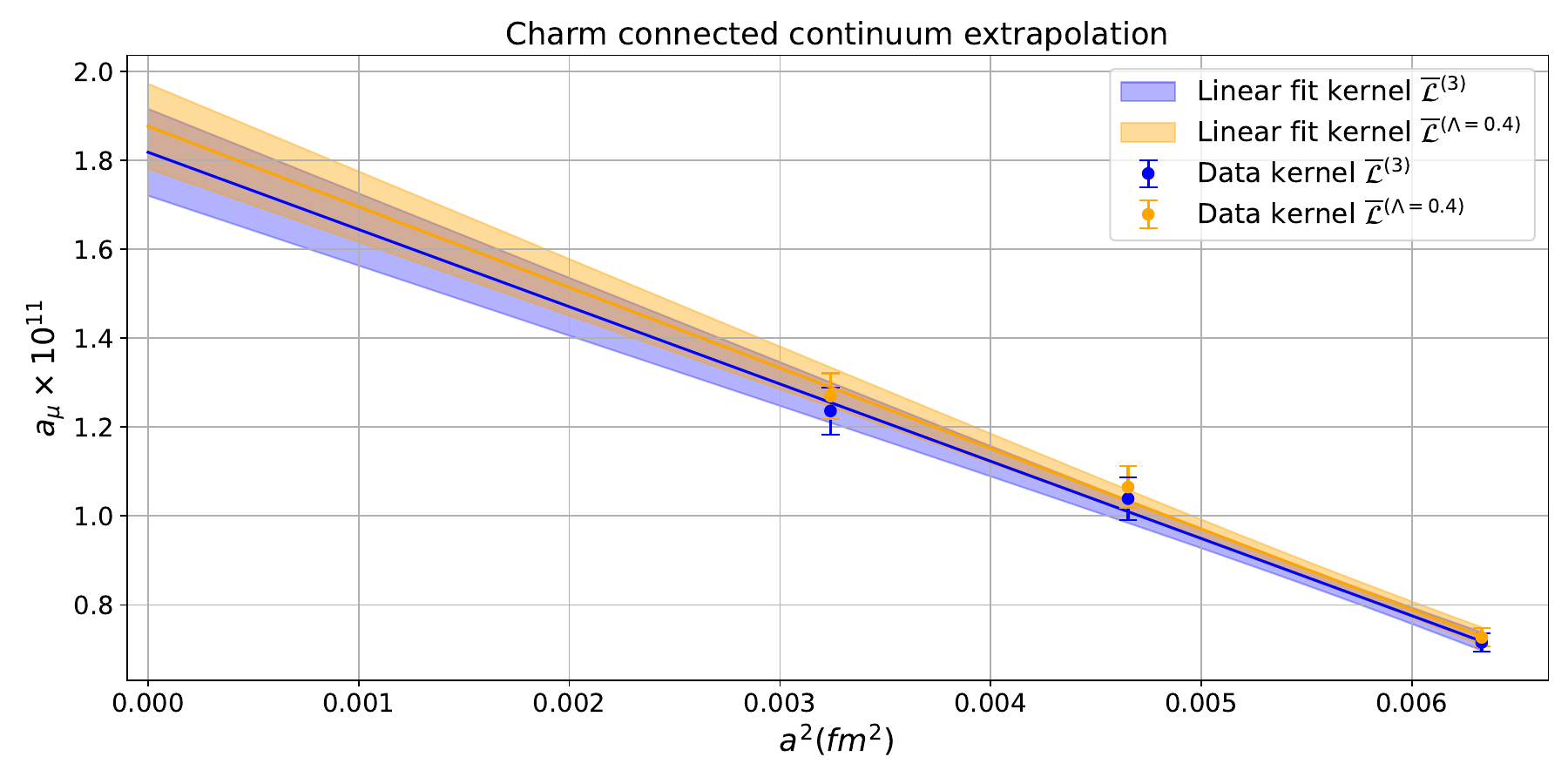}
    \caption{Continuum extrapolation of the charm-quark connected contribution.}
    \label{fig:continuum-charm-connected}
\end{figure}

In Figure \ref{fig:charm-kernels}  we show our preliminary results for the charm-quark connected contribution $a_\mu^\text{HLbL,c,conn}$ on the ensembles cB64, cC80, and cD96 for kernels $\overline{\mathcal{L}}^{(\Lambda=0.4)}$ (upper two plots) and $\overline{\mathcal{L}}^{(3)}$ (lower two plots). In both cases, the left plot depicts the integrand $f(|y|)$ as a function of the distance $|y|$ and the right plot the partially integrated $a_\mu(|y|_\text{max})$ as a function of the cut-off distance $|y|_\text{max}$. In Figure \ref{fig:continuum-charm-connected} we show a preliminary continuum extrapolation of our resuts for $a_\mu^\text{HLbL,c,conn}$. It uses a linear fit in $a^2$, which describes the data well, giving a $\chi^2/\text{d.o.f.} = 0.61$ for kernel $\overline{\mathcal{L}}^{(\Lambda=0.4)}$ and $\chi^2/\text{d.o.f.} = 0.51$ for kernel $\overline{\mathcal{L}}^{(3)}$. The statistical error is controlled at the level $\sim5\%$ for both kernels, based on measurements using 100 configurations for cB64 and 31 configurations for cC80 and cD96 spread out uniformly across each ensemble. 
An estimate of the systematic error stemming from the continuum extrapolation and
tuning of the charm-quark mass needs to be done, before any meaningful comparisons with existing results \cite{Chao:2022xzg} are possible. 

\begin{figure}[h!]
    \centering
    \begin{subfigure}[b]{\textwidth}
        \centering
        \includegraphics[width=\textwidth]{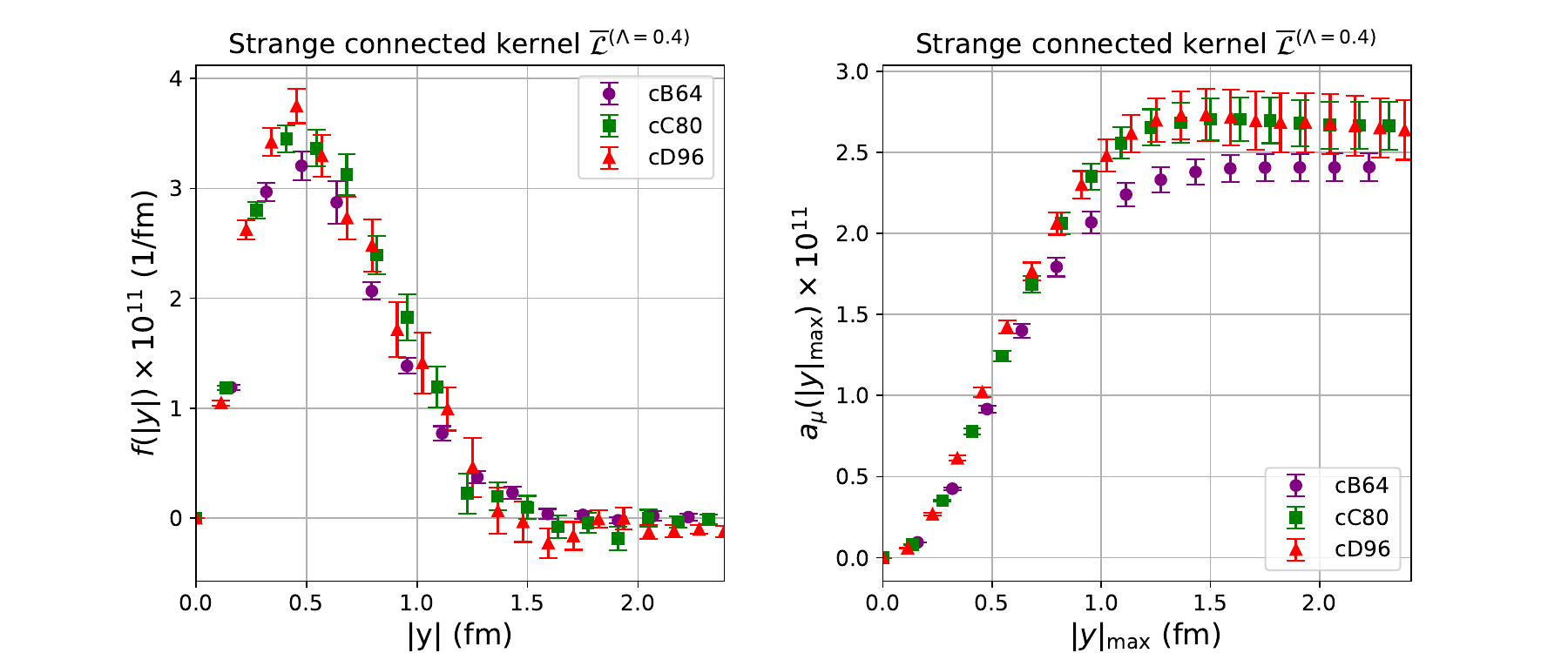}
        \label{fig:strange-k4}
    \end{subfigure}
    
    \begin{subfigure}[b]{\textwidth}
        \centering
        \includegraphics[width=\textwidth]{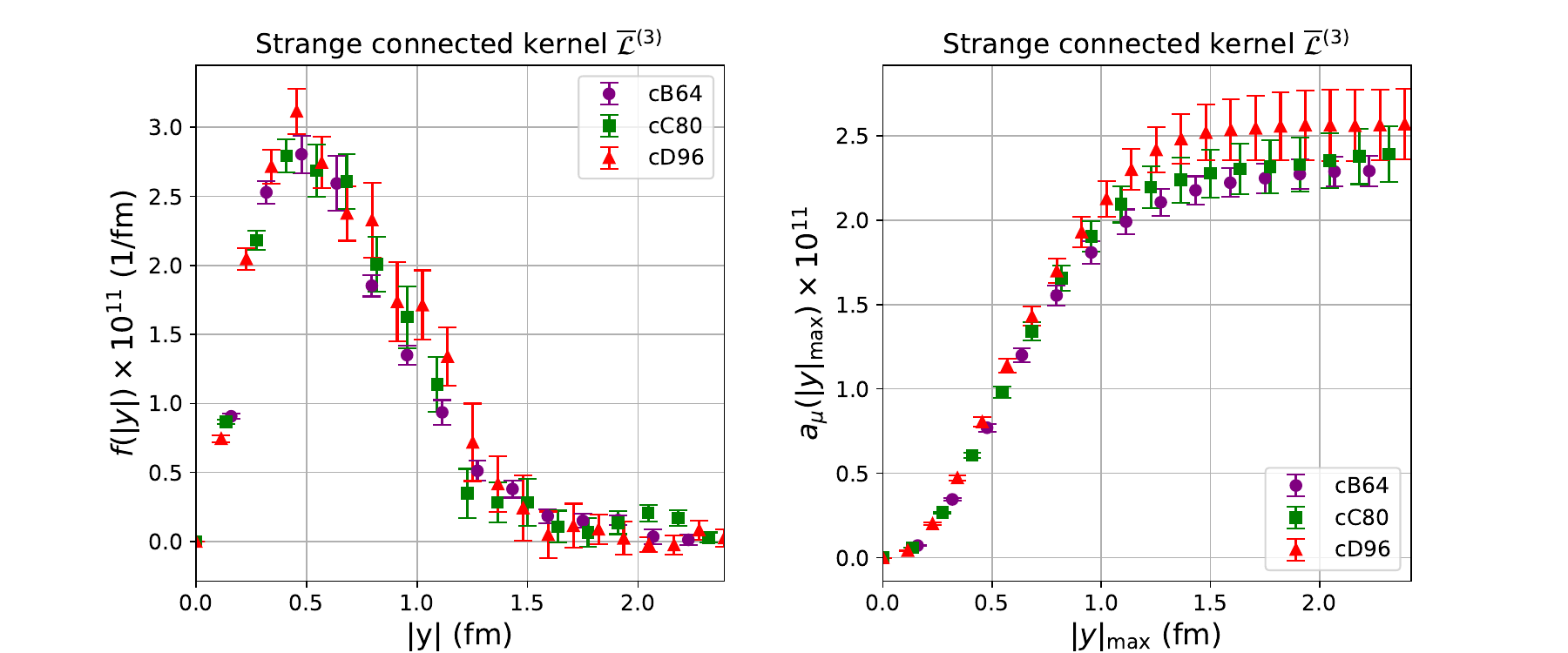}
        \label{fig:strange-k3}
    \end{subfigure}
    
    \caption{Preliminary results for the strange-quark connected contribution on the ensembles cB64, cC80, and cD96 for kernels $\overline{\mathcal{L}}^{(\Lambda=0.4)}$ (upper two plots) and $\overline{\mathcal{L}}^{(3)}$ (lower two plots). In both cases, the left plot depicts the integrand $f(|y|)$ as a function of the distance $|y|$ and the right plot the partially integrated $a_\mu(|y|_\text{max})$ as a function of the cut-off distance $|y|_\text{max}$.}
    \label{fig:strange-kernels}
\end{figure}

\begin{figure}[h!]
    \centering
    \hspace*{-0.5cm} 
    \includegraphics[width=0.8\textwidth]{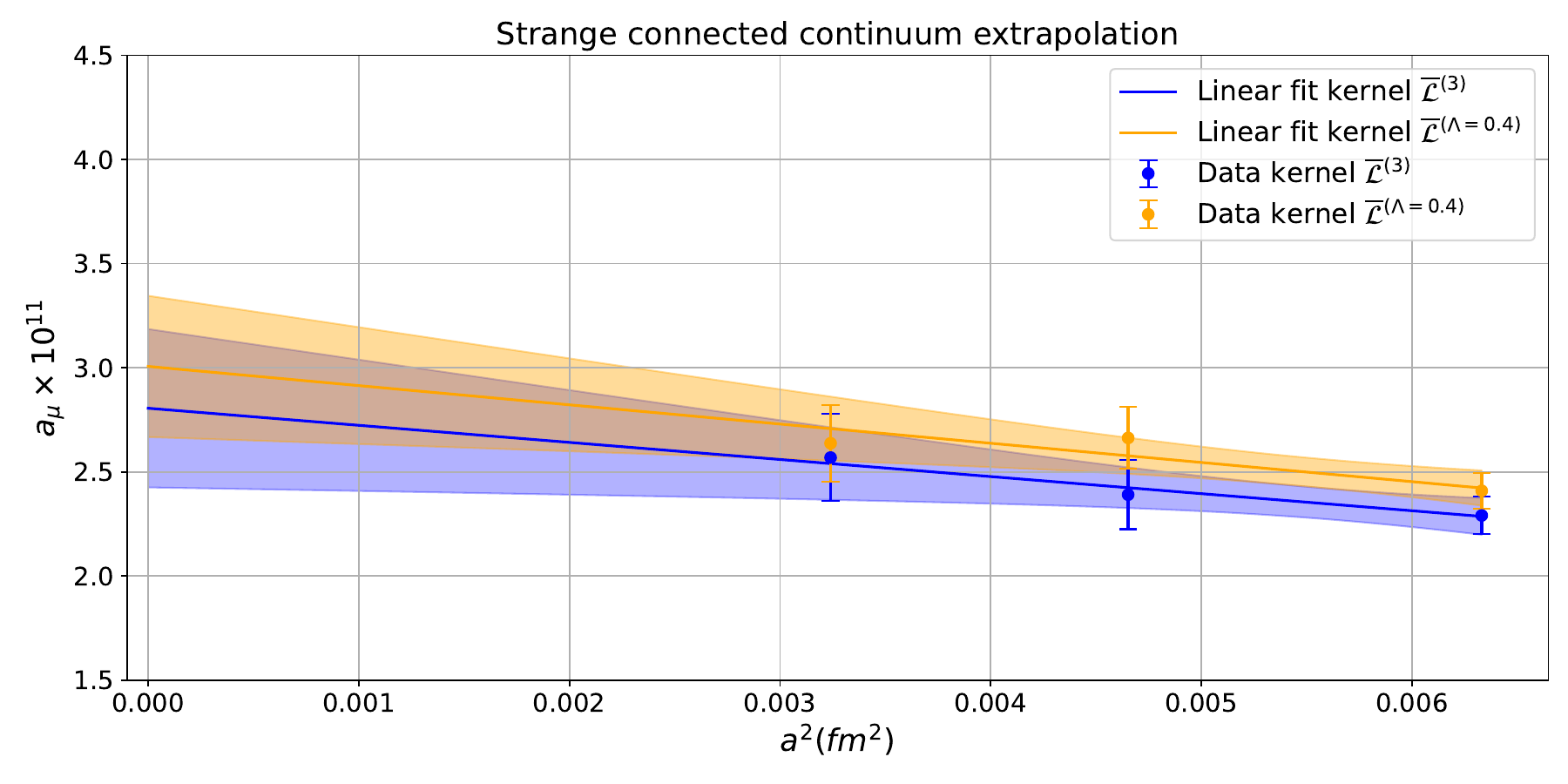}
    \caption{Continuum extrapolation of the strange-quark connected contribution.}
    \label{fig:continuum-strange-connected}
\end{figure}

We have likewise evaluated the strange-quark connected contribution on all three ensembles. These results can be seen in Figure \ref{fig:strange-kernels} and \ref{fig:continuum-strange-connected}. We have again performed a continuum extrapolation with a linear fit in $a^2$, which seems to describe the data well for both kernels,
with a $\chi^2/\text{d.o.f.} = 0.50$ for kernel $\overline{\mathcal{L}}^{(\Lambda=0.4)}$ and $\chi^2/\text{d.o.f.} = 0.06$ for kernel $\overline{\mathcal{L}}^{(3)}$. In this case, the lattice artefacts are smaller compared to the charm-quark connected case and the continuum extrapolation is flatter. We used 48 configurations for cB64, 54 configurations for cC80 and 30 configurations for cD96, spread out uniformly across each ensemble. The statistical error is $\sim12\%$ for both kernels, while the systematic error still needs to be estimated (in progress).

\noindent

\noindent

\label{subsec:strange_connected}
\noindent

\noindent

\subsection{Light-quark connected contribution}
\label{subsec:light_connected}

\begin{figure}[h!]
    \centering
    \includegraphics[width=\textwidth]{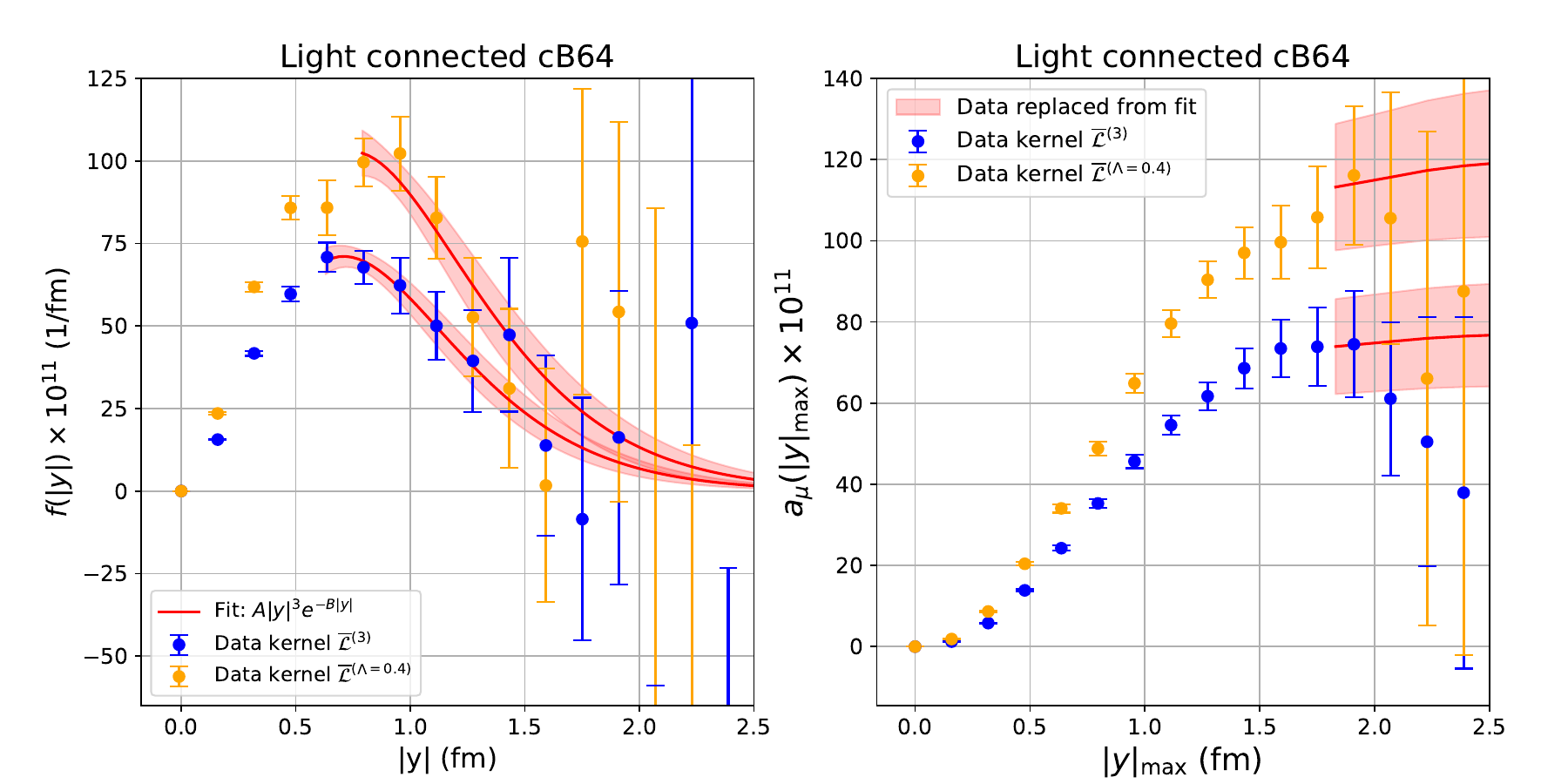}
    \caption{
    Preliminary results for the light-quark connected contribution on ensemble cB64.
    The left plot depicts the integrand $f(|y|)$ as a function of the distance $|y|$ with a fit of the form $A~|y|^3~e^{-B|y|}$ for $|y| \gtrsim 0.6$ fm as a red band. The right plot shows the partially integrated $a_\mu(|y|_\text{max})$ as a function of the cut-off distance $|y|_\text{max}$ with the effect of replacing the data by the fit result for $|y| \gtrsim 1.9$ fm. 
    }
    \label{fig:light-connected-k3-k4}
\end{figure}

In this subsection we present the preliminary results of the light-quark connected contribution, evaluated on the coarsest ensemble (cB64), with lattice spacing $a \sim 0.08$ fm and with 789 configurations. These results can be seen in Figure \ref{fig:light-connected-k3-k4}. As a first observation, the quality of the signal is good up to $|y| \lesssim 1.5$ fm, however the data quickly become noisy after $|y| \gtrsim 1.5$ fm. One way to tackle this problem is to replace the data in the tail with data from the neutral-pion exchange, which is expected to yield the dominant contribution to $a_{\mu}^{\text{HLbL}}$ at large distances. We aim to use the data from our previous work on the $\pi^0$ transition form factor reported in Ref.~\cite{ExtendedTwistedMass:2023hin}.
Alternatively, one can parameterize the pole contribution---or in general the single-resonance exchange---at large distances via a fit to the form $A~|y|^3~e^{-B|y|}$, as suggested in Ref.~\cite{Chao:2021tvp}.
In this preliminary investigation
we choose to fit the integrand from $|y| \geq 0.79$ fm  for kernel $\overline{\mathcal{L}}^{(\Lambda=0.4)}$ and from $|y| \geq 0.63$ fm for $\overline{\mathcal{L}}^{(3)}$, yielding $\chi^2/\text{d.o.f.} = 0.70$ 
and $\chi^2/\text{d.o.f.} = 0.74$,
respectively. The right panel of Figure \ref{fig:light-connected-k3-k4} depicts the effect of replacing the data after $|y| \geq 1.9$ fm using the results of these fits, and it can be seen how the result for $a_{\mu}$ is stabilized in the long-distance limit $|y|_\text{max} \rightarrow \infty$.
In the future, we aim to compare results from a variety of fit ranges versus the direct $\pi^0$-pole contribution computed on the same ensemble. The systematic error associated with our replacement of large-distance data with data from the fits is not included in the present analysis.

\noindent

\subsection{Light-quark disconnected $(2+2)$ contribution}
\label{subsec:light-light_2+2}

\begin{figure}[h!]
    \centering
    \includegraphics[width=\textwidth]{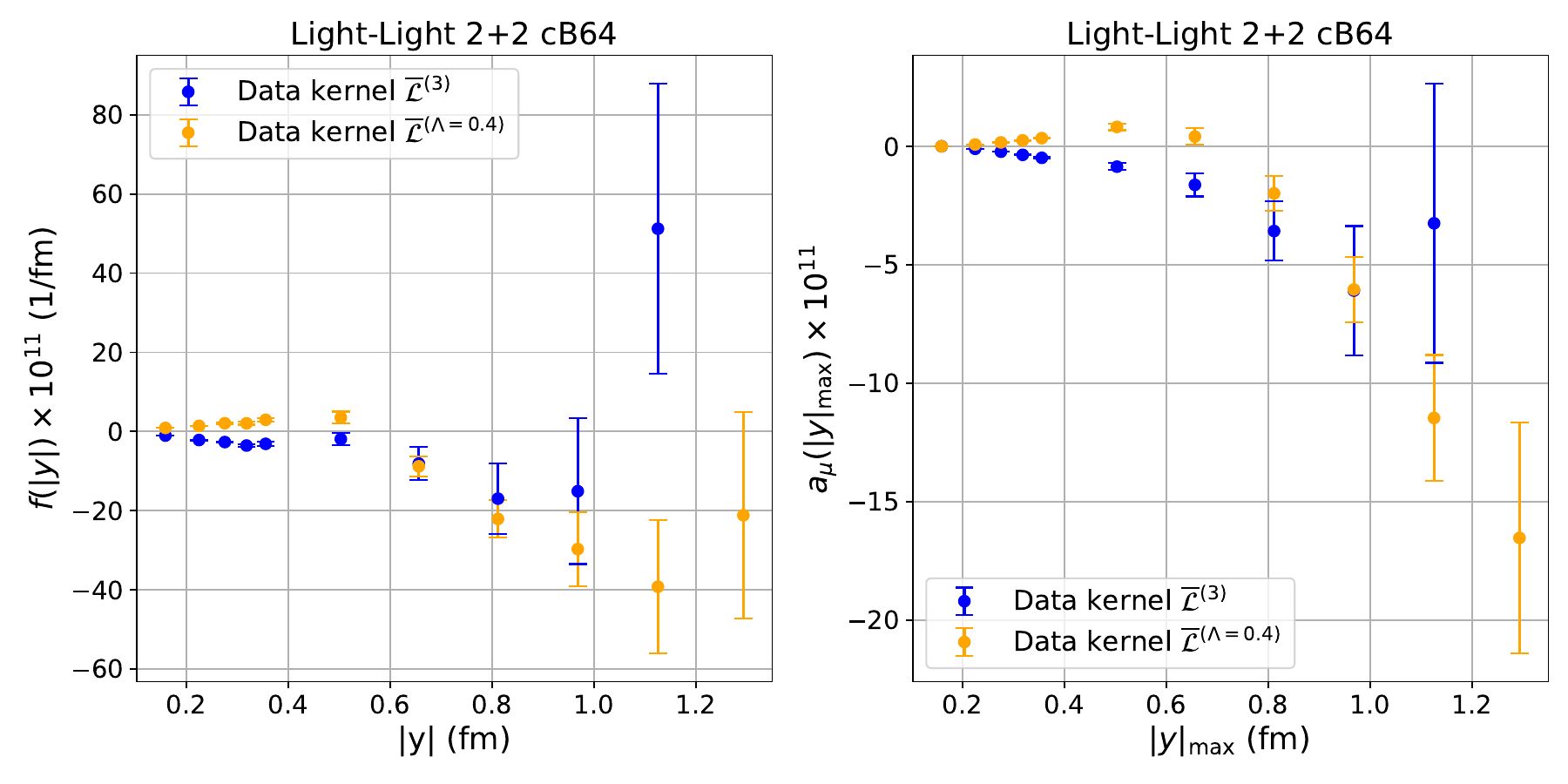}
    \caption{Preliminary results for the light-quark disconnected $2+2$ contribution on ensemble cB64. The left plot depicts the integrand $f(|y|)$ as a function of the distance $|y|$ and the right plot the partially integrated $a_\mu(|y|_\text{max})$ as a function of the cut-off distance $|y|_\text{max}$.}
    \label{fig:light-light-cB}
\end{figure}

The light-quark disconnected $(2+2)$ contribution is the leading disconnected contribution and here we present our findings for the coarsest ensemble (cB64) with lattice spacing $a \sim 0.08$ fm. From Figure \ref{fig:light-light-cB}, one can see that the statistical error on $a_\mu$ for kernel $\overline{\mathcal{L}}^{(\Lambda=0.4)}$ is below $30\%$ up to $1.2$ fm. We have also evaluated the contribution using the kernel $\overline{\mathcal{L}}^{(3)}$ for a comparison of the lattice artefacts,
but only using approximately four times fewer statistics, leading to larger uncertainties. We plan to increase statistics for both kernels, to produce data at larger distances, and to analyse the light-strange and strange-strange quark disconnected $(2+2)$ contributions. In order to control the statistical noise in the disconnected quantities for larger distances
we expect to follow a similar approach as for the connected case, replacing the long-distance tail of $f(|y|)$ by the $\pi^0$- and $\eta$-pole data that can be obtained from our computations of the corresponding transition form factors in Refs.~\cite{ExtendedTwistedMass:2023hin, ExtendedTwistedMass:2022ofm}.  

\noindent

\section{Conclusions and outlook}

We have presented preliminary results for the  HLbL contribution to the muon anomalous magnetic moment from our lattice-QCD computation using $N_f = 2+1+1$ twisted-mass fermions on gauge ensembles at the physical point produced by the ETMC collaboration.
For the charm- and strange-quark connected contributions, we have so far obtained results
at three values of the lattice spacing and in the continuum limit, however, various systematic errors still need to be estimated for a final result. 

For the light-quark connected contribution, we have so far performed measurements
on the ensemble cB64 with lattice spacing $a \sim 0.08$ fm, with sufficient statistics to effectively determine the integrand
up to $|y| \sim 1.5$ fm. However,
to achieve the goal of $O(10\%)$ precision for $a_\mu^\text{HLbL}$,
a way to treat the noisy data at large distances has to be implemented.
So far, we have performed an exponential fit to the data using an ansatz inspired by the $\pi^0$-pole exchange, which significantly improves the statistical precision of the integrand at large distances. The next steps are
to estimate systematic uncertainties from this process using multiple fit ranges and our direct evaluation of the $\pi^0$-pole contribution,
as well as the extension of our calculations to the other two ensembles cC80 and cD96 with finer lattice spacings. 

The light-quark disconnected $(2+2)$ contribution has so far been evaluated on the coarsest ensemble cB64 with lattice spacing $a \sim 0.08$ fm. We get a good signal up to $|y| \sim 1.1$ fm for the kernel with the higher statistics, $\overline{\mathcal{L}}^{(\Lambda=0.4)}$. We plan to increase the precision and produce data at larger distances for both kernels. We also plan to calculate the light-strange and strange-strange disconnected $(2+2)$ contributions and extend our calculations to
the ensembles with finer lattice spacings. 

{\bf Note added:} While finalizing these proceedings, new results appeared from the BMW collaboration \cite{Fodor:2024jyn} which indicate a significantly higher continuum-limit value for $a_\mu^{\text{HLBL, c, conn}}$ at the physical point as compared to our result. The source of the discrepancy needs to be investigated. 

\acknowledgments

We thank En-Hung Chao and Antoine G\'erardin for helpful discussions. This work is supported by the Swiss National Science Foundation (SNSF) through grant No.200020$\_$208222. MP acknowledges support from the Sino-German collaborative research center CRC 110. We gratefully acknowledge computing time granted on Piz Daint at Centro Svizzero di Calcolo Scientifico (CSCS) under the project ID s1197. We gratefully acknowledge the Gauss Centre for Supercomputing e.V. (www.gauss-centre.eu) for funding this project through computing time on the GCS supercomputer JUWELS Booster at the Jülich Supercomputing Centre. Evaluations of the QED kernel were possible by using the KQED library~\cite{Asmussen:2022oql}. Ensemble production and measurements for this analysis made use
of tmLQCD~\cite{Jansen:2009xp, Deuzeman:2013xaa, Abdel-Rehim:2013wba, Kostrzewa:2022hsv} and QUDA~\cite{Clark:2009wm, Babich:2011np, Clark:2016rdz}. The figures were produced using matplotlib~\cite{4160265}.

\bibliographystyle{JHEP}
\bibliography{main} 

\end{document}